# From Local Earthquake Nowcasting to Natural Time Forecasting: A Simple Do-It-Yourself (DIY) Method

John B Rundle[1,2,3], Ian Baughman[1], Andrea Donnellan[4,3], Lisa Grant[5], Geoffrey Fox[6]

[1] University of California, Davis, CA
[2] Santa Fe Institute, Santa Fe, NM
[3] Jet Propulsion Laboratory, Pasadena, CA
[4] Purdue University, West Lafayette, IN
[5] University of California, Irvine, CA
[6] University of Virginia, Charlottesville, VA

## Abstract

Previous papers have outlined nowcasting methods to track the current state of earthquake hazard using only observed seismic catalogs. The basis for one of these methods, the "counting method", is the Gutenberg-Richter (GR) magnitude-frequency relation. The GR relation states that for every large earthquake of magnitude greater than $M_T$, there are on average $N_{GR}$ small earthquakes of magnitude $M_S$. In this paper we use this basic relation, combined with the Receiver Operating Characteristic (ROC) formalism from machine learning, to compute the probability of a large earthquake. The probability is conditioned on the number of small earthquakes $n(t)$ that have occurred since the last large earthquake. We work in natural time, which is defined as the count of small earthquakes between large earthquakes. We do not need to assume a probability model, which is a major advantage. Instead, the probability is computed as the Positive Predictive Value (PPV) associated with the ROC curve. We find that the PPV following the last large earthquake initially decreases as more small earthquakes occur, indicating the property of temporal clustering of large earthquakes as is observed. As the number of small earthquakes continues to accumulate, the PPV subsequently begins to increase. Eventually a point is reached beyond which the rate of increase becomes much larger and more dramatic. Here we describe and illustrate the method by applying it to a local region around Los Angeles, California, following the January 17, 1994 magnitude $M$6.7 Northridge earthquake.

## Key Points

- Earthquake nowcasting tracks the current risk of a large earthquake in local regions by counting small earthquakes
- The method can be extended to future-looking earthquake forecasting using standard machine learning methods
- A major advantage is that the forecast probability curve is determined directly from the data itself, rather than being assumed

## Plain Language Summary

Forecasting and predicting the location and time of major earthquakes are long-sought goals. Unlike weather forecasting, the data needed for detailed and precise earthquake forecasting will always be incomplete. The state of tectonic stress and the strength of faults, which are the data needed for such predictions, cannot be observed by any means currently known. As a result, we must rely on indirect methods. What can be observed is data represented by earthquake catalogs, which include the time, location, and magnitude of the events. In past works, we have shown that the current state of the local regions can be observed by counting the number of small earthquakes since the last large earthquake in a region. We have called this method earthquake nowcasting. In the present paper, we use the statistics of these small events, combined with signal detection methods from machine learning, to compute the probability of future large earthquakes given the current state of the fault system. We then illustrate these procedures by application to the region around Los Angeles, CA following the January 17, 1994 magnitude M6.7 Northridge earthquake. Major advantages of the proposed method are its basic simplicity, and the fact that there are no unknown parameters that must be assumed or set by arbitrary means.

## Introduction

**Background.** Earthquake nowcasting is the estimation of the current state of a seismically active region. The term nowcasting is used in the same sense as for weather and economic nowcasting (e.g., Rundle et al., 2016; Rundle et al., 2021a), the determination of the current state of a system in the recent past, current time, and the near future. Methods to produce earthquake nowcasts have been the subject of previous papers, a selection of which are: (Rundle et al., 2016; 2018; 2019a; 2019b; 2020; 2021a,b; 2022a,b; 2024; Pasari and Mehta, 2018; Pasari, 2019; Pasari, 2020; Pasari and Sharma, 2020; Chouliaras, 2009; Chouliaras et al., 2023; Perez-Oregon, 2020). In general, these methods take three forms:

- A "counting" method using counts of small earthquakes of a small completeness magnitude $M_S$ to track the chance of future large earthquakes (Rundle et al., 2016) having a target magnitude $M_T$, $M_T \gg M_S$.
- A "filter" method, in which the monthly rate of small earthquakes is filtered and optimized to produce a time series that gradually increases prior to large earthquakes, and decreases sharply after them (Rundle et al., 2022a).
- An "eigenpattern" method in which a regional time series is constructed by expanding the space time patterns of small earthquakes in a series of the eigenvectors of a correlation matrix (Rundle et al., 2022b).

The basis for the counting method is the Gutenberg-Richter magnitude frequency relation. The filter method produces a time series that strongly resembles the hypothesized buildup and release of tectonic stress, along with a spatial probability density function indicating probable locations of the future large earthquakes. The filter is optimized by a machine learning method based on the skill measure of the associated Receiver Operating Characteristic (ROC), leading to the computation of the precision, or Positive Predictive Value (PPV), which in turn is the probability of a future large earthquake. The eigenpattern method produces a time series that closely resembles the time series produced by the filter method.

In this paper, we exploit the counting method, together with the ROC and PPV to produce local earthquake forecasts in seismically active circular regions around points of interest. We apply the method to a circular region of radius 125 km around the city of Los Angeles, CA, USA. A major advantage of this method is that the probability of a future large earthquake is determined directly from the catalog data, with no arbitrary assumptions.

The forecast method developed here is in natural time, counts of small earthquakes. To transform this method to clock or calendar time, a mapping is needed to associate earthquake counts with clock/calendar time. While we briefly address this topic in the discussion, we defer a more detailed consideration to a future paper.

**Natural Time and a Principle of Statistical Equivalence.** In this paper we work in natural time (Varotsos 2001, 2011, 2014; Holliday et al, 206b; Rundle et al., 2021a). Natural time is defined as the count of small earthquakes of magnitude $M_S$ that have occurred since the last large "target" earthquake having magnitude $M_T$. Given the Gutenberg-Richter relation, every large earthquake of magnitude greater than $M_T$ is associated with, on average, $N_{GR}$ small earthquakes of magnitude $M_S$ as explained below.

In the counting method, we consider a seismically active circular region centered on a point of interest. The circular region is embedded in a larger region with a substantially more target earthquakes than the circular region. In addition, the larger region is selected to have the same Gutenberg-Richter statistics (i.e., *b*-value) as the circular region, implying that both have the same statistics. We consider this similarity to represent a principle of "statistical equivalence", so that we can use the much larger ensemble of target earthquakes in the region to create the natural time forecast of the few target earthquakes in the circle. Note that there must have been at least one earthquake withing the circle with magnitude greater than $M_T$.

Over the same long calendar time interval, the long term occurrence rate of small earthquakes per unit calendar time in the large region, $R_R$, will be larger than the long term rate of small earthquakes in the circle, $R_C$: $R_R \gg R_C$. There are a correspondingly a larger number of target earthquakes as well, which constitute the ensemble of events to consider.

In the forecast method describe below, we apply the ROC method in the large region to compute the PPV in the circle in natural time. We begin by defining a future number (natural time) of small earthquakes $\Delta N_C$ in the circular region that will be used for the forecast. For purposes of reference, we can define an average calendar time scale by:

$$\frac{\Delta N_C}{R_C} \equiv T_{Scale} \qquad (1)$$

This calendar time scale **should not** be regarded as a future forecast time interval, because the rate of accumulation of small earthquakes is not constant, and can in fact be highly variable as is well known (e.g., Rundle et al., 2022a), a point that is further discussed below.

## Basic Method

We consider an active circular region embedded in a larger active surrounding region. We use the GR relation in the form (e.g., Boore, 1989):

$$N_{GR} = 10^{b\{M_T - M_S\}} \qquad (2)$$

where $M_T$ is the magnitude of the large target earthquake of interest, and $M_S$ is the small (completeness) magnitude, and $b$ is the GR $b$-value.

We build a time series in the larger region as described below. The time series consists of a group of earthquake "cycles" between the target earthquakes. The length of cycle $i$, $L(C_i)$, is then the number of small earthquakes between target earthquakes bounding the beginning and end of cycle $i$.

To construct and analyze the time series, we follow these initial steps:

1. A large region containing a statistically significant number of earthquakes (>~ 20) of the target magnitude $M_T$ is defined, surrounding the circular region of radius $R$. The circular region must contain at least one previous earthquake of magnitude $M_T$. The size of the region is adjusted so that the GR statistics of the region match the GR statistics of the circular region.

2. Since it is known (e.g., Gardner and Knopoff, 1974) that the interval statistics of earthquakes are generally Poisson distributed, we define an "accumulation function" $A_F$ as:

$$A_F = 1 - \exp\left(-\frac{n}{N_{GR}}\right) \qquad (3)$$

Here $n$ is the number of small earthquakes (magnitude $M_S$) that have occurred since the last large earthquake. Thus we work in "natural time", as described above. In this method, the current value of the accumulation function is equivalent to the Earthquake Potential State (EPS) as defined in (Rundle et al., (2016; 2018; 2019a; 2019b; 2020; 2021a,b; 2022; 2024). Note specifically that we do not have any information on $n(t)$, i.e., how the number accumulates with time.

3. A time series is constructed in the large region in which the value of the time series is given by the accumulation function (3) in the interval between target earthquakes.

4. We define "cycles of activity", where the number of cycles corresponds to the number of target earthquakes (minus 1), as will be seen in the figures shown in the application below. When a target earthquake occurs, the accumulation value of the time series is reset to zero, and the next cycle begins.

5. A natural time (not calendar time) forecast window for the future number of small earthquakes, $\Delta N_C$, is selected for the circle.

**Receiver Operating Characteristic (ROC) Analysis**

An ROC analysis is conducted on the time series in the large region to build the ROC curve, as described in Rundle et al. (2022). But instead of using the clock or calendar time until the next target earthquake, we use natural time. We determine whether the length $L(C)$ of a given earthquake cycle has more small earthquakes than the current number $n$ since the last target earthquake. If it does, then we further determine 1) whether n is less than or greater than a threshold value $n_{th}$; and 2) whether the total number of small earthquakes in a given cycle lies between the current number, $n$, and the current number plus the expected future number, $n+\Delta N_C$. For threshold values $n_{Th}$, we use the number of small earthquakes between $n = 0$ and the maximum cycle length $L_{max}(C)$. All values of $n_{Th}$ are considered successively to classify the time series.

Working in natural time, the implementation of the ROC analysis follows the same procedure as in Rundle et al. (2022) to define True Positive (TP), True Negative (TN), False Positive(FP) and False Negative (FN). Note that TP, TN, FP, and FN will be functions of the threshold value $n_{Th}$. For each cycle, we classify the small earthquakes as follows:

- If $n \geq n_{Th}$, and the next target earthquake **does occur** during when $n+\Delta N_C$ more earthquakes have occurred, $n$ is classified as True Positive, TP($n_{Th}$).

- If $n \geq n_{Th}$, and the next target earthquake **does not occur** during when $n+\Delta N_C$ more earthquakes have occurred, $n$ is classified as False Positive, FP($n_{Th}$).

- If $n < n_{Th}$, and the next target earthquake **does occur** during when $n+\Delta N_C$ more earthquakes have occurred, $n$ is classified as False Negative, FN($n_{Th}$)

- If $n < n_{Th}$, and the next target earthquake **does not occur** during when $n+\Delta N_C$ more earthquakes have occurred, $n$ is classified as True Negative, TN($n_{Th}$).

The process is then repeated for the ensemble of cycles, and the values are aggregated into the categories TP($n_{Th}$), FP($n_{Th}$), TN($n_{Th}$), FN($n_{Th}$). Note that the sum over all thresholds of these categories must equal the total number of small earthquakes in the time series.

We now define (Mandrekar, 2010; Powers, 2011) the hit rate or True Positive Rate (TPR), the false alarm rate, or False Positive Rate (FPR), and the precision, or Positive Predictive Value (PPV):

$$TPR = \frac{TP}{TP + FN} \quad FPR = \frac{FP}{FP + TN} \quad PPV = \frac{TP}{TP + FP} \quad (4)$$

The ROC diagram is a plot of the *TPR* vs. the *FPR*. The Area Under Curve (AUC), or "skill" is a measure of the value of the method. A skill value of 0.8 or larger is generally deemed to be excellent (Powers, 2011). The precision, or PPV, is then the forecast of target earthquakes as a function of the threshold value $n_{Th}$.

## Example: Application to Los Angeles Region of Southern California

We apply this analysis to events in a circle of radius 125 km around Los Angeles, California. We consider target earthquakes $M_T \geq 6.0$, and a catalog beginning in 1970, with small earthquakes $M_T > M > M_S = 3.49$. In the python code referenced in the Open Access section below, we include a search in a square surrounding region for side lengths in which the GR *b*-values of region and circle are close in value.

Using a simple optimization, we found that a box dimension of 7° x 7° in latitude-longitude centered on Los Angeles gave a *b*-value for $b_{region}$ = 0.934 ± 0.008, close to the value for $b_{circle}$ = 0.925 ± 0.019. *b*-values were determined for the magnitude range [3.75, 6.5]. Note that since 1970, 2 earthquakes of the minimum target magnitude $M_T$ = 6.0 occurred within the circle, the February 9, M6.6 1971 San Fernando earthquake, and the January 17, M6.7 Northridge earthquake.

Figure 1a shows the seismicity in the large surrounding region. In Figure 1b the time series is constructed and plotted using the accumulation function (2). Again the accumulation value represents the Earthquake Potential State (Rundle et al, 2016, d202a).

Figure 2 is a diagram with the ROC curves at left, and PPV curves at right. The top PPV curve is PPV plotted as a function of the accumulation function values $A_F$ with a red dot indicating the current value as of 9/23/2025. The bottom curve is PPV as a function of the accumulating small earthquake number *n*, plotted through 9/23/2025, with the red dot terminating the current value of *n* = 447.

In Figure 2a, the ROC diagram is computed before any small earthquakes have occurred. In Figure 2b, the ROC diagram is computed for the number of small earthquakes, 447, that have occurred since the 1994 Northridge earthquake. We choose a natural time interval in the circle $\Delta N_C$ = 44 small events (corresponding to $T_{Scale}$ = 3 years). Note that 44 small events is approximately the number of events that occurred during the first 3 hours following the 1994 Northridge earthquake. Recently it has taken more than 3 years, since February 2022 until the present, to accumulate 44 small earthquakes within the Northridge circle. Thus $T_{Scale}$ should not be regarded as a fixed forecast time scale as described above.

The cyan curves in the ROC diagram are the curves for 50 random forecasts. These random forecasts were computed by building 50 random time series using a bootstrap random sampling algorithm with replacement. The black diagonal line is the mean of these 50 random curves, and the dotted lines represent one standard deviation from the mean. The diagonal line thus represents a forecast with no information. The skill score is the area under the ROC curve, which is AUC = Skill = 0.46 for Figure 2a, indicating basically random skill. In Figure 2b, the Skill = 0.88, indicating much better skill. A score greater than 0.8 has been classified to represent "excellent" predictive skill (Mandrekar, 2010), a value greater than 0.9 is "oustanding". We also define a "skill index", which is defined as 100% * |(skill - 0.5)/0.5|. Skill index has this form since a skill lower than 0.5 represents skill for forecasting a "non-event" (Rundle et al., 2024).

An important caveat is that these ROC diagrams are constructed with the requirement that the cycles in the ensemble used must have at least as many small events as the current count, which in Figure 2a is 1, and for Figure 2b is 447. Thus the number of usable cycles for Figure 2a is 43, many of which represent "noise cycles", and for Figure 2b is 8, which are the much longer-lived cycles, and presumably more "deterministic" and relevant cycles.

On Figure 2c, we plot the Precision, or Positive Predictive Value (PPV) as a function of the accumulation value ($A_F$) for the current earthquake potential state in the circle in which there are currently 447 small earthquakes since the 1994 M6.7 Northridge, California earthquake. Note the Gutenberg-Richter "wall" at the right hand side, corresponding to the longest cycle in the ensemble, with a length of 846 small earthquakes.

On Figure 2d, we plot the PPV as a function of the accumulating number of small earthquakes that have occurred since the Northridge earthquake. Two distinct regimes or phases can be seen. The first regime is an initial probability of a subsequent target earthquake $M \geq 6.0$ of ~17.4%, corresponding to enhanced mainshock clustering. This initial probability then decreases to ~13.5% when ~120 small earthquakes have occurred. Following that initial period, a second period of re-loading leading up to the next $M \geq 6.0$ target earthquake in the circle can be seen. This second period begins at small earthquake number ~120 and increases as more small earthquakes occur, to a present value of about 22.5% at the current number 447 of small earthquakes in the circle.

Figure 3a shows the current state of the target earthquake potential in the circle, representing the final slide of the movie that can be produced with the python code. In this figure, the green bars represent the binned number of small earthquakes in the 43 earthquake cycles in the large region, a histogram of the total of 43 cycles. The stair-stepping red curve is the cumulative distribution function (CDF), derived from the histogram. The magenta curves close to the red curve represent the $1\sigma$ standard deviation from the histogram, again using a bootstrap method. The dashed blue line is the accumulation function, computed from equation (2).

In Figure 3b, the red "thermometer" has the same value as the CDF (for easy reference), representing the EPS value. The green "anti-thermometer" = 100% - EPS value (the red thermometer), and is the survival distribution function (SDF) value. In Figure 3c we plot the total number of small earthquakes following the 1994 Northridge earthquake up to the present time, color coded with more recent earthquakes in hotter colors. Large circle marks the epicenter of the Northridge mainshock.

## Discussion

The foregoing results can be computed with the Python code that accompanies this paper. The URL for the code is listed in the "Open Research" section below. The code can be used to produce a movie of the development of the nowcast/forecast. Figure 3 is the final slide in a movie. The code includes a "quickstart" README file that has instructions on how to set up and run the code.

A remaining problem is to map the natural time forecast into a calendar/clock time forecast. To develop this idea, we need a mapping between count increments $\Delta N_C$ and calendar/clock time increments $\Delta t_C$ at time $t$ following the last target earthquake. For aftershocks, this mapping is essentially the Omori relation describing aftershock decay. This relation implies that for constant $\Delta t_C$, $\Delta N_C$ decreases inversely with a power of $t$ as time increases. At longer ("non-aftershock") times following the last target earthquake, the mapping may be different. We will defer more detailed consideration of this idea to a future publication.

**Acknowledgements**. Research by JBR was supported in part by a grant from the Southern California Earthquake Center grant #SCON-00007927 to UC Davis, and by the John LaBrecque fund, a generous gift from John LaBrecque to the University of California, Davis. SCEC contribution number 14988. The authors would also like to acknowledge an informative conversation with Jeanne Hardebeck.

**Open Research.** Python code that can be used to reproduce the results of this paper can be found at the Zenodo site: https://doi.org/10.5281/zenodo.17290440

**Data.** Data for this paper was downloaded from the USGS earthquake catalog for California, and are freely available there. An included method in the Python code mentioned above can be used to download these data for analysis.

**Figure Captions**

**Figure 1.** a) Map of seismicity used in the example, the optimal region of 7º latitude x 7º longitude centered on Los Angeles, CA. Earthquakes having magnitudes larger than M6 are large red circles as shown in the figure key. The circle of radius 125 km is shown in blue. b) Time series of the accumulation of small earthquakes as a function of time.

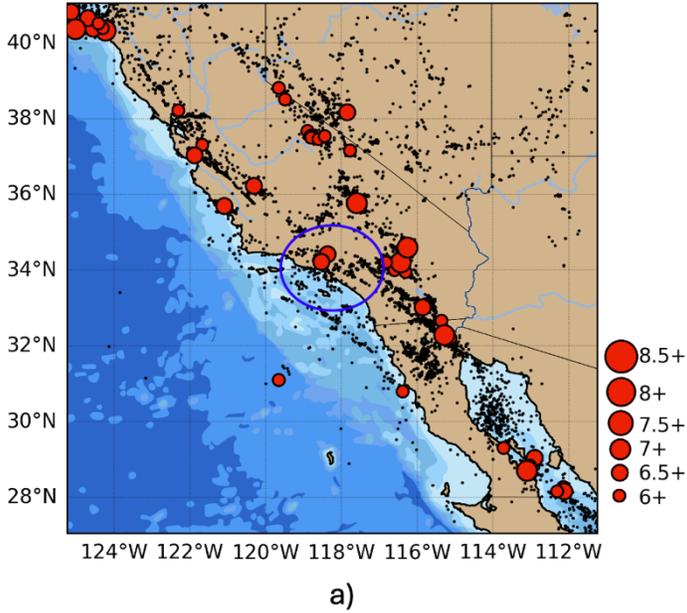
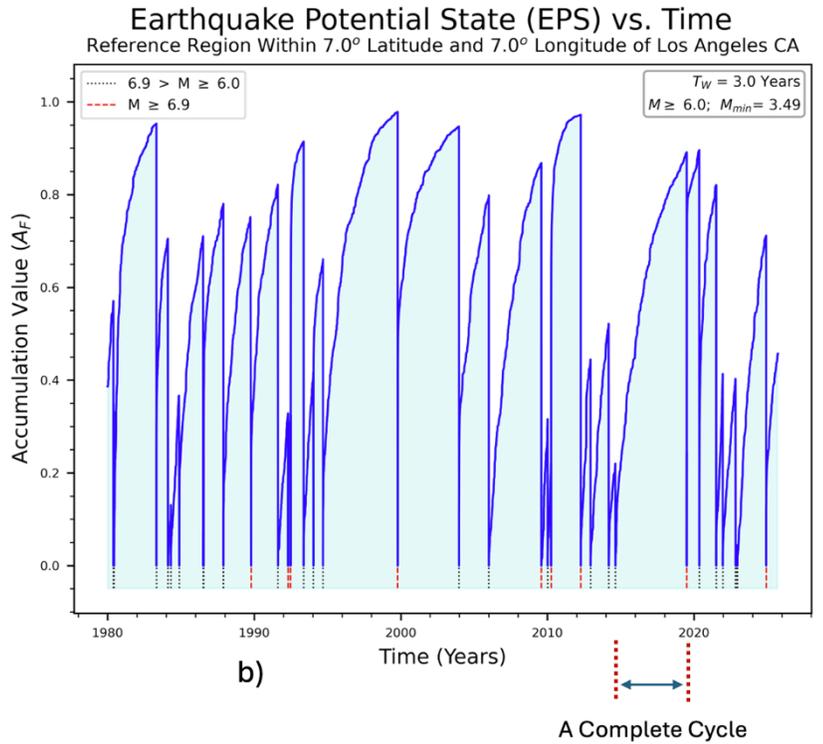

*From Earthquake Nowcasting to Earthquake Forecasting*

**Figure 2.** a) Red curve is the Receiver Operating Characteristic (ROC) diagram as described in the text, a plot of True Positive Rate (TPR, hit rate), vs. False Positive Rate (FPR, false alarm rate) just after the last large earthquake when no small earthquakes have yet occurred. The number of useful cycles at this stage is 43 complete cycles since 1980, of the 45 cycles since 1970. Cyan curves represent ROC curves for 50 random forecasts. b) ROC diagram for the present day when 447 small earthquakes have occurred since the Northridge earthquake, so that only 8 of the 43 cycles can be used. c) Plot of the Precision, or Positive Predictive Value (PPV) as a function of the accumulation value ($A_F$) for the current earthquake potential state in the circle. d) Plot of the PPV as a function of number of small earthquakes that have occurred since the Northridge earthquake. Two distinct regimes or phases can be seen: 1) An initial probability of a subsequent $M$6.0 target earthquake of 17.4%, corresponding to enhanced mainshock clustering, which then decreases to 13.5% until ~120 small earthquakes have occurred. 2) A second period of re-loading leading up to the next $M$6.0 target earthquake in the circle, beginning at small earthquake ~120 and increasing as more small earthquakes occur, to a present value of about 22.5%.

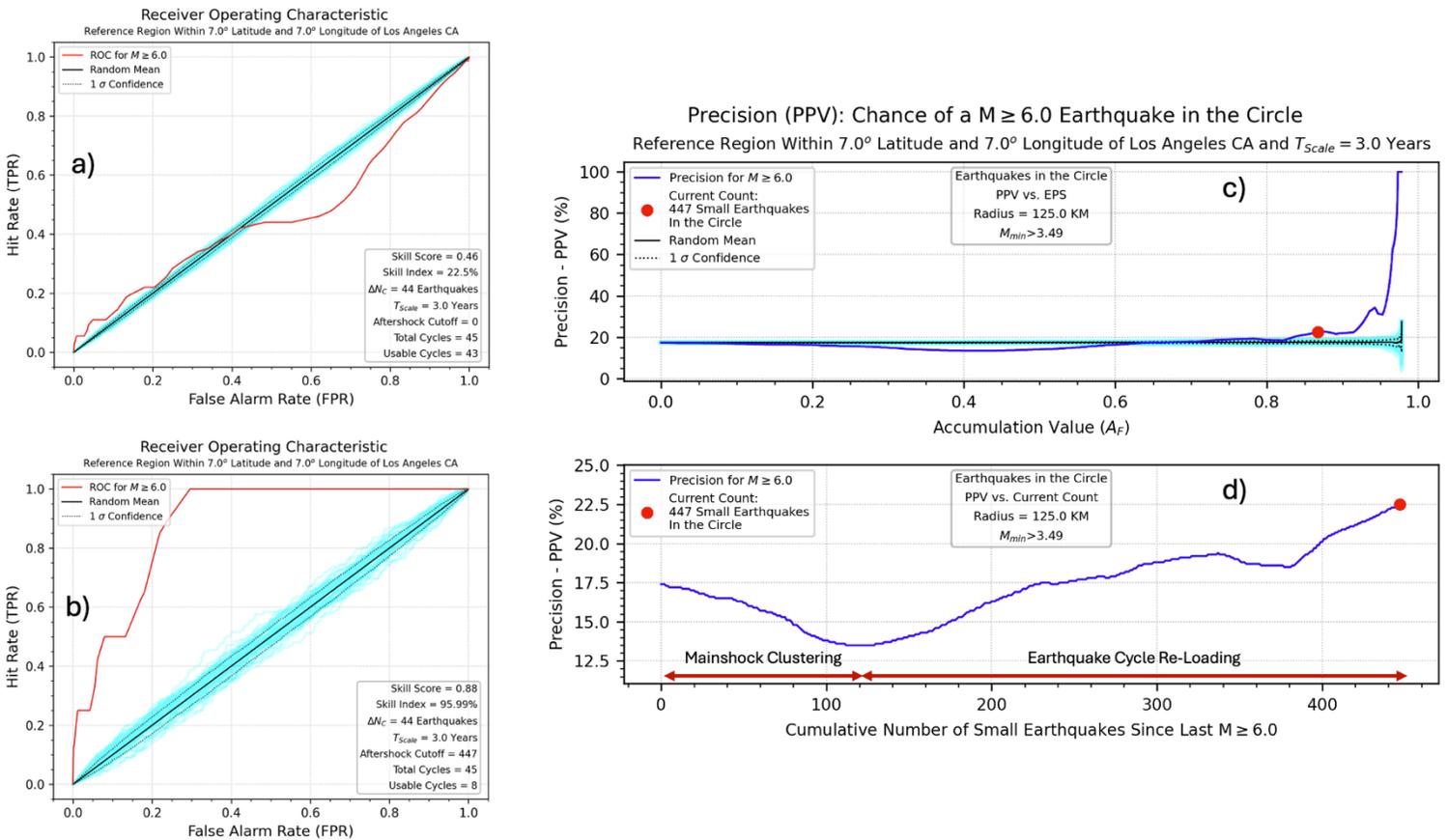

**Figure 3.** Current state of the target earthquake potential in the circle, representing the final slide of the movie that can be produced with the python code. a) Green bars represent the histogram of small earthquakes in the 45 complete earthquake cycles in the large region since 1970. The Cumulative Distribution Function (CDF) is derived from the histogram. The magenta curves close to the red curve represent the $1\sigma$ deviation from the histogram, again using a bootstrap method. The dashed blue line is the accumulation function, computed from equation (2). b) The red "thermometer" has the same value as the CDF (for easy reference), representing the EPS value. The green "anti-thermometer" = 100% - EPS value (the red thermometer), and is the Survival Distribution Function (SDF) value. c) Total of small earthquakes following the 1994 Northridge earthquake up to the present time, 10/7/2025, color coded with more recent earthquakes in hotter colors. Large circle marks the epicenter of the Northridge mainshock.

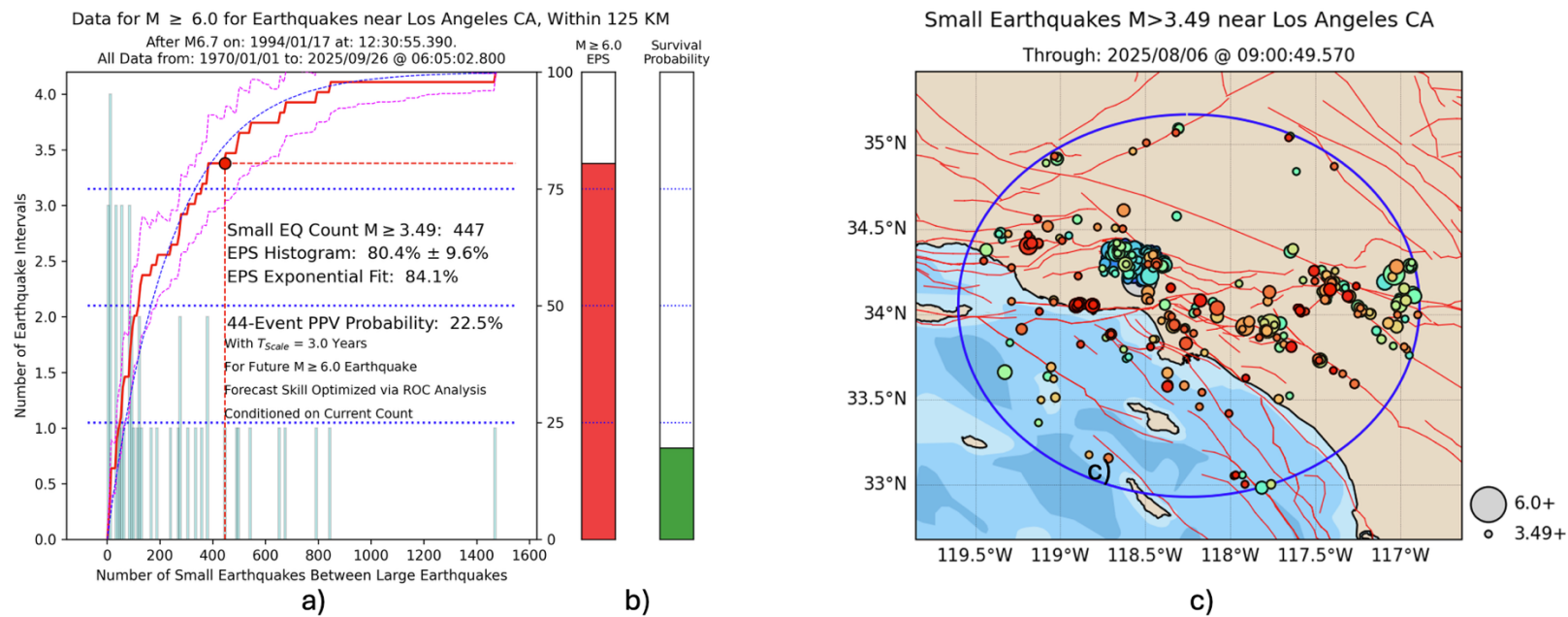